\documentclass[a4paper,12pt,nohead]{article}

\usepackage{cite}
\usepackage{setspace}
\usepackage{graphicx}
\usepackage{hyperref}
\usepackage[lmargin=3cm,rmargin=1cm,tmargin=2cm,bmargin=2cm]{geometry}

\newcommand{\rmd}{\mathrm{d}}
\newcommand\figref[1]{Fig.~\ref{#1}}

\newcommand{\subfigfff}{0.95\textwidth}

\title{The consentaneous model of the financial markets exhibiting spurious nature of long-range memory}

\author{V. Gontis, A. Kononovicius}

\date{Institute of Theoretical Physics and Astronomy, Vilnius University}

\begin{document}

\maketitle

\begin{abstract}
It is widely accepted that there is strong persistence in the volatility of financial time series. The origin of the observed persistence, or long-range memory, is still an open problem as the observed phenomenon could be a spurious effect. Earlier we have proposed the consentaneous model of the financial markets based on the non-linear stochastic differential equations. The consentaneous model successfully reproduces empirical probability and power spectral densities of volatility. This approach  is qualitatively different from models built using fractional Brownian motion. In this contribution we investigate burst and inter-burst duration statistics of volatility in the financial markets employing the consentaneous model. Our analysis provides an evidence that empirical statistical properties of burst and inter-burst duration can be explained by non-linear stochastic differential equations driving the volatility in the financial markets. This serves as an strong argument that long-range memory in finance can have spurious nature.
\end{abstract}

\section{Introduction}

We have to acknowledge that current understanding of the financial fluctuations and the nature of microscopic market interactions remains limited and ambiguous
\cite{Farmer2012EPJ,Shiller2014AER,Kirman2014MD}. This imposes a natural limits on estimating risk in the financial markets and is directly related to the complex market dynamics involved 
\cite{Kwapien2012PhysRep,Bouchaud2004Cambridge,Sornette2004Princeton}.  Statistical physics 
is a useful tool to deal with complexity 
in the financial markets \cite{Karsai2012NIH,Chakraborti2011RQUF1,Gabaix2009AR} as a greater insight is achieved using advanced methods of empirical analysis \cite{Campbell1996Princeton,Mantegna2000Cambridge,Drozdz2010NJP,Farmer2012EPJ}.

A major problem in the modeling in finance is related to the double stochastic nature of the fluctuations in the real markets. First of all there is a wide consensus on the need to model the behavioral opinion dynamics of the traders in the financial markets \cite{Brock2001RES,Diks2005JEDC,Franke2012JEDC,Lux2012JEDC,Goldbaum2014JEBO,He2015JEF,Jang2015CE} and many models proposed are able to explain the fat tails and the volatility clustering. Usually these models describe oversimplified stochastic behavior with a limited number of the parameters and statistically adjusted values, which are not universal for the various definitions of the financial variables and various other statistical properties. Seeking for the heuristic model with universal parameters it is necessary to combine endogenous (agent-based) fluctuations with exogenous noise arising from the information or the order flow. Starting from the phenomenological stochastic modeling of return \cite{Gontis2010Intech} we have proposed the consentaneous agent-based and stochastic model \cite{Gontis2014PlosOne} (further in the text we refer to this model as the consentaneous model), which reproduces probability density function (PDF) and power spectral density (PSD) of the absolute return in the financial markets. The endogenous dynamics of volatility in this model is based on  the stochastic differential equations (SDEs) derived for the infinite number of agents with global herding interactions. The time series of the high-frequency return in this model are generated by combining endogenous volatility with exogenous Gaussian fluctuations. First of all the consentaneous model with the same set of parameters reproduces PDF and PSD of absolute return for various assets and different markets. The statistical properties of the consentaneous model scale in the same way as the empirical data does for different return time-scale.

Later it was shown that the consentaneous model is able to explain  various statistical properties of the high volatility return intervals extensively studied before in
\cite{Yamasaki2005PNAS,Wang2006PhysRevE,Wang2008PhysRevE,Bunde2011EPL,Bunde2014PRE}.
Our empirical study \cite{Gontis2016PhysA} using the consentaneous model across a wide range of time-scales from one minute to one month has demonstrated that proposed concept of financial fluctuations allows to understand statistics of volatility return intervals. In that study it was shown that for the sufficiently high values of the threshold the PDF of volatility return intervals has universal scaling with the prevailing power-law exponent $3/2$. This inspired us for the further  empirical study of burst and inter-burst duration PDFs for the time series of trading activity and absolute return (see \cite{Gontis2017PhysA}), which are usually considered to have the long-range memory. The power-law with exponent $3/2$ in burst and inter-burst duration PDF probably means that Markov processes might be behind the stochastic dynamics of financial markets. 

Here we employ the consentaneous model to demonstrate how various noises overlap and coexist finally resulting in the observed statistical properties of the burst and inter-burst duration. Being based on a Markov processes the consentaneous model helps us to explain the spurious nature of the long-range memory in the financial markets. In Section 2 we shortly discuss the structure of the consentaneous model, in Section 3 we compare empirical and model statistical properties of the burst and inter-burst duration, in Section 4 we analyze the effect of the various noises included in the consentaneous model on the PDFs of the burst and inter-burst duration. Finally we conclude the results presented in this paper.

\section{Consentaneous model of the financial markets}
We have already used the consentaneous model \cite{Gontis2014PlosOne} to reproduce and explain the statistical properties of the volatility return intervals \cite{Gontis2016PhysA} and to argue for the necessity of the exogenous noise in the modeling of financial markets \cite{Gontis2016APPA}. Here we describe the model in a very general terms seeking to reveal its relevance to the problem of the long-range memory. As was demonstrated in \cite{Gontis2017PhysA,Gontis2017Entropy} the burst and inter-burst duration PDFs help to discriminate between two different origins of the observed long-range memory. The fundamental power-law with exponent $3/2$ indicates about one-dimensional Markov processes in the origin of fluctuations when deviations from this law might be related with true long-range memory processes such as fBm. Our preliminary empirical analysis of the FOREX data \cite{Gontis2017PhysA} confirmed the presence of power-law with exponent $3/2$ for the time series of trading activity and absolute returns. Here we demonstrate how the consentaneous model can be used to show that the financial markets might be driven by the long-term stochastic process described by non-linear SDE. 

First lets recall that time series of return $r_{\delta}(t)=lnP(t)-lnP(t-\delta)$ related with market price $P(t)$ in sufficiently short time period $\delta$, of one minute order, is defined in the model as \cite{Gontis2014PlosOne}
\begin{equation}
r_{\delta}(t) = \sigma(t) \omega(t),
\label{eq:return}
\end{equation}
here $\omega(t)$ denotes a Gaussian exogenous noise, related to the order flow fluctuations, and $\sigma(t)$ is the slowly varying endogenous volatility (assumed to be almost constant for the time windows of width $\delta$). Volatility being result of agent dynamics itself is a double stochastic process defined by ratio $y(t)=\frac{1-n_f}{n_f}$ of chartists $1-n_f$ and fundamentalists $n_f$ as well by the mood of speculative traders $\xi(t)$,
\begin{equation}
\sigma(t) = b_0(t)(1+ a_0 \vert y(t) \xi(t) \vert),
\label{eq:defvolatil}
\end{equation}
here the empirical parameter $a_0$ determines the impact of the agent dynamics on the observed time series. We account for the daily seasonality observed in the real data by introducing a periodic time dependence \cite{Gontis2014PlosOne} of volatility 
\begin{equation}
b_0(t)=\exp [-(\{t \mathrm{mod} 1 \}
  -0.5)^2/w^2]+0.5,
\label{eq:b0}
\end{equation}
where $w=0.25$ quantifies the width of intra-day pattern.  The most important part of this approach is related to the stochastic processes $n_f(t)$, $y(t)$ and $\xi(t)$, which can be modeled using ordinary SDEs and they are thus Markov processes. Nevertheless, even in this case Markov processes $\sigma(t)$ and $y(t)$ exhibit the long-range memory properties, such as power spectral density $S(f) \sim f^{-\beta}$ with $\beta \approx 1$. Lets recall the SDEs defining these stochastic processes in the consentaneous model
\begin{eqnarray}
& \rmd n_f = \frac{(1-n_f) \varepsilon_{cf} - n_f
    \varepsilon_{fc}}{\tau(n_f)} \rmd t + \sqrt{\frac{2 n_f
      (1-n_f)}{\tau(n_f)}} \rmd W_{f} , \label{eq:nftau}\\ 
& \rmd \xi = - \frac{2 h_{cc} \varepsilon_{cc} \xi}{\tau(n_f)} \rmd t +
  \sqrt{\frac{2 h_{cc} (1-\xi^2)}{\tau(n_f)}} \rmd W_{\xi} , \label{eq:xitau} 
\end{eqnarray}
where the inter-trade time $\tau(n_f)$ takes the form
\begin{equation}
\frac{1}{\tau(n_f)}= \left( 1 + a_{\tau} \left| \frac{1-n_f}{n_f}
\right| \right)^{\alpha}, \label{eq:taunfxi} 
\end{equation}
with empirical parameter $a_{\tau}$. Equations (\ref{eq:nftau}) and (\ref{eq:xitau}) describe long term stochastic dynamics of the fundamentalists $n_f$ and the stochastic dynamics of chartists' mood $\xi$ (which is $h_{cc}$ times quicker). The both processes are defined by the global herding interactions among the traders making choice between fundamental and speculative trading behavior in Eq. (\ref{eq:nftau}) and between optimism and pessimism in Eq. (\ref{eq:xitau}). All parameters of the idiosyncratic agent transitions $\varepsilon_{cf}$ (chartists-fundamentalists), $\varepsilon_{fc}$ (fundamentalists-chartists) and $\varepsilon_{cc}$ (optimists-pessimists or pessimists-optimists) are normalized here by the herding parameter $h$, defining the main slow time scale of agent dynamics between the chartists and the fundamentalists. $W_f$ and $W_{\xi}$ are independent standard Wiener processes, parameter $\alpha=2$ is defined by the empirical analyses of the trading activity and the return \cite{Rak2013APPB,Gabaix2003Nature,Farmer2004QF,Gabaix2006QJE}.

Eqs. (\ref{eq:nftau})-(\ref{eq:taunfxi}) serve as a macroscopic description of the agent-based (endogenous) dynamics and together with Eqs. (\ref{eq:return}) and (\ref{eq:defvolatil}) they constitute a complete set of equations behind the consentaneous model. 

The key stochastic variable in this model is the ratio $y(t)=\frac{1-n_f}{n_f}$ defined by SDE (\ref{eq:nftau}) with the trading activity given by (\ref{eq:taunfxi}). This can be written in the form of knotty equation, but let us write an almost equivalent non-linear SDE for $y$ applicable for the symmetric $\tau$  introduced in \cite{Gontis2017Entropy}
\begin{equation}
\rmd y=\left(\varepsilon_1 y^{-\alpha} + (2-\varepsilon_2)y^{1-\alpha}\right)(y+1)^{2\alpha+1} \rmd t + \sqrt{2y^{1-\alpha}}(y+1)^{\alpha+1} \rmd W,
\label{eq:SDEy}
\end{equation}
It is worth to not here that the SDE for $y$ belongs to the class of equations generating stochastic variables  with power-law properties in the first and second order statistics, see few other papers on the subject \cite{Kaulakys2005PhysRevE,Ruseckas2011PhysRevE,Ruseckas2014JStatMech}. It is very important for this contribution that despite few other noises present in the consentaneous model main statistical properties of return time series may arise from the general class of  non-linear SDEs.
 
\section{Comparison between the model and the empirical PDFs of the burst and inter-burst duration}

In \cite{Gontis2017PhysA} we have made empirical evaluation of the PDFs for the burst and inter-burst duration in the filtered time series of the FOREX absolute return and trading activity. It was expected that the deviations from the power-law with exponent $3/2$ would give us an indication of the true long-range memory as theory predicts for the processes such as fBm. Such test is not unconditional as a real processes such as the volatility in the financial markets is more complicated than one-dimensional stochastic processes. Thus this and any other empirical tests should be evaluated while taking into account the  structure of the stochastic processes present in the financial markets. From our point of view the consentaneous model provides such structure and with success reproduces the other statistical properties of the empirical time series. Here we compare the statistical properties of the burst and inter-burst duration in the filtered FOREX time series of the absolute return. The burst and inter-burst durations are considered as two distinct threshold passage events -- first describes return of the signal to the threshold from above and is denoted as $T$, while the second describes return to the threshold from below and is denoted as $\theta$, see \cite{Gontis2017PhysA} for the details. 

\begin{figure}
\centering
\includegraphics[width=\subfigfff]{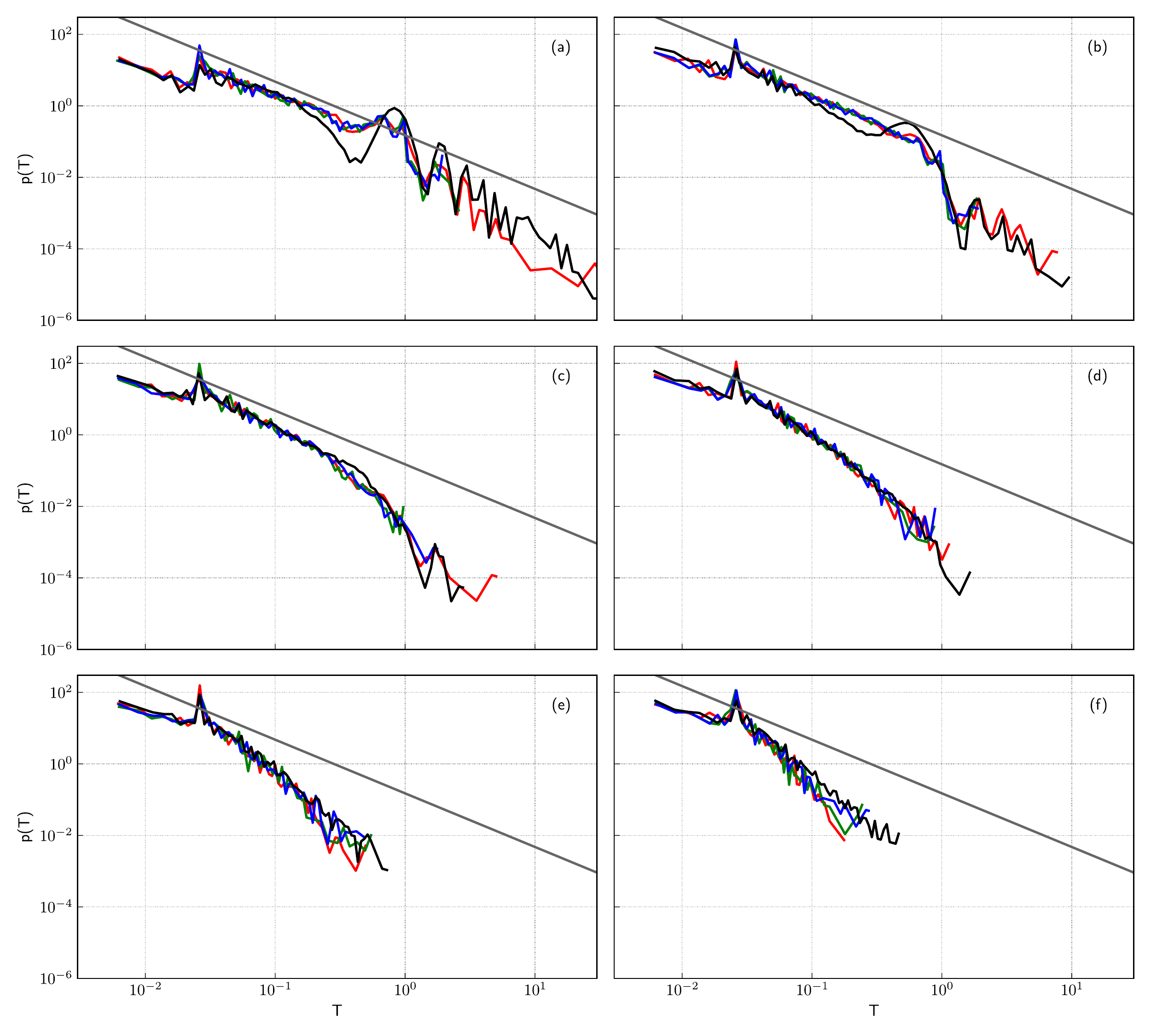}
\caption{\label{fig-1} PDFs of the burst duration $T$ for the empirical and the model time series of the absolute return. EUR/USD exchange series (red curve), XAG/USD series (green curve), XAU/USD series (blue curve), model series (black curve). Values of the threshold $q$ were set as follows: (a) - $0.3$; (b) - $0.5$; (c) - $0.8$; (d) - $1.3$; (e) - $2$; (f) - $3$. The straight gray curves are shown to guide the eye showing a power-law with the exponent $3/2$.}
\end{figure}

\begin{figure}
\centering
\includegraphics[width=\subfigfff]{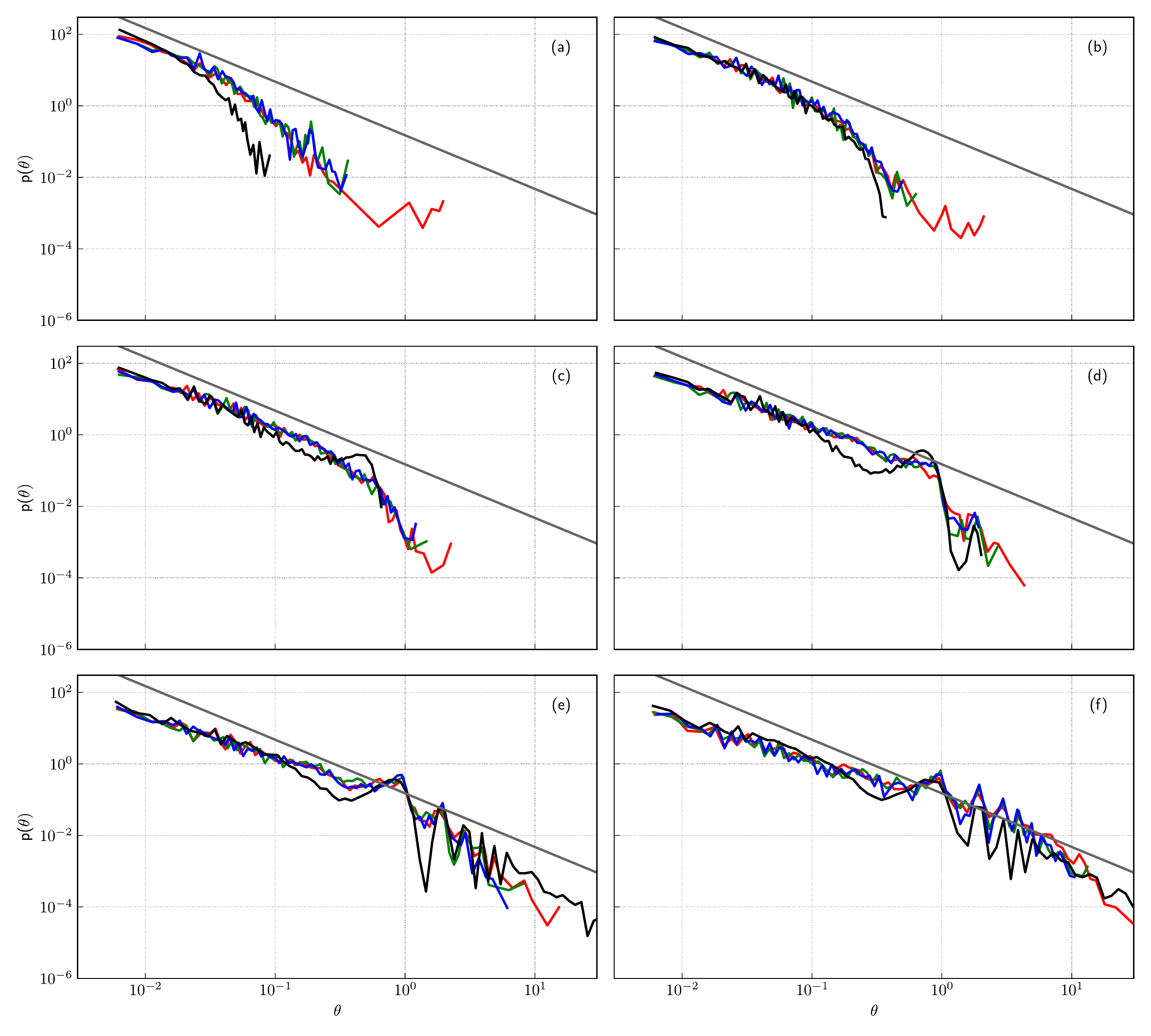}
\caption{\label{fig-2} PDFs of the inter-burst duration $\theta$ for the empirical and the model time series of the absolute return. EUR/USD exchange series (red curve), XAG/USD series (green curve), XAU/USD series (blue curve), model series (black curve). Values of the threshold $q$ are the same as in Fig. \ref{fig-1}. The straight gray curve are shown to guide the eye showing a power-law with the exponent $3/2$.}
\end{figure}

We use here the same set of model parameters as in \cite{Gontis2016PhysA}: $\varepsilon_{cf}=1.1$ and $\varepsilon_{fc}=3$, $\varepsilon_{cc}=3$, $H=1000$, $a_0=1$, $a_{\tau}=0.7$, $\alpha=2$ and $h=0.3 \cdot 10^{-8}
s^{-1}$.  All the parameter values are kept constant
throughout the analysis that follows. For the convenience time is given in days and the basic time period of return definition in FOREX series $\delta=1/390$ is equivalent to $221$ seconds.  We filter the exogenous noise $\omega$, Eq. (\ref{eq:return}), in empirical and model time series by using the standard deviation filter with time window of $10 \times \delta$. The PDFs of $T$ for different values of threshold $q$: (a) - $q=0.3$; (b) - $q= 0.5$; (c) - $q=0.8$; (d) - $q=1.3$; (e) - $q=2$; (f) - $q=3$; which are measured in the standard deviations of the time series, are given in \figref{fig-1} and PDFs of $\theta$ in \figref{fig-2}. Note that this threshold set will be used through out all subsequent figures in this contribution.

Lets recall that the consentaneous model was built to reproduce PDF and PSD of the absolute return time series. It does it with high precision and for $\delta=1/390$ empirical and model data give us, for example, values of exponents for PSD  $\beta_1=1.4$ and $\beta_2=0.5$, see \cite{Gontis2017PhysA} for details. The value of Hurst parameter, defined from the standard relation with the exponent of PSD $\beta=2H+1$, is $H=(\beta_1-1)/2=0.2$. Thus the exponent of corresponding burst and inter-burst duration distribution $2-H=1.8$ should be expected, \cite{Ding1995fbm}. This implies a meaningful deviation from the expected $3/2$ law, which should be observable in the PDFs of the burst and inter-burst duration, if the considered time series were one-dimensional stochastic Markov processes.

One can observe in \figref{fig-1} and \figref{fig-2} a very good coincidence of the $T$ and $\theta$ PDFs for a few assets traded on FOREX and the consentaneous model given for wide choice of the threshold values. The model PDF for different values of the threshold $q$ scales precisely in the same way as empirical PDFs do. Considerable deviation of the model from empirical data is observed only for PDF of $\theta$, when the threshold is very low ($q=0.3$). This is the case, when interplay of fluctuations related to exogenous noise, daily seasonality and speculative trading is the most important and complicated. 

In the previous empirical study \cite{Gontis2017PhysA} we have concluded that these results serve as an argument that dynamics of the volatility in the financial systems probably is based on Markov processes. So good correspondence of the consentaneous model with the empirical data adds confidence  to the more detailed investigation of the spurious memory in the financial markets using the proposed consentaneous model. Certainly, this model allows us to evaluate the contribution of all fluctuations to the burst and inter-bust PDFs included and accounted there. In the following section we will decompose the consentaneous model looking more deeply into behavior of the burst and inter-burst duration.

\section{Contribution of the various noises to the PDFs of burst and inter-burst duration}

\begin{figure}
\centering
\includegraphics[width=\subfigfff]{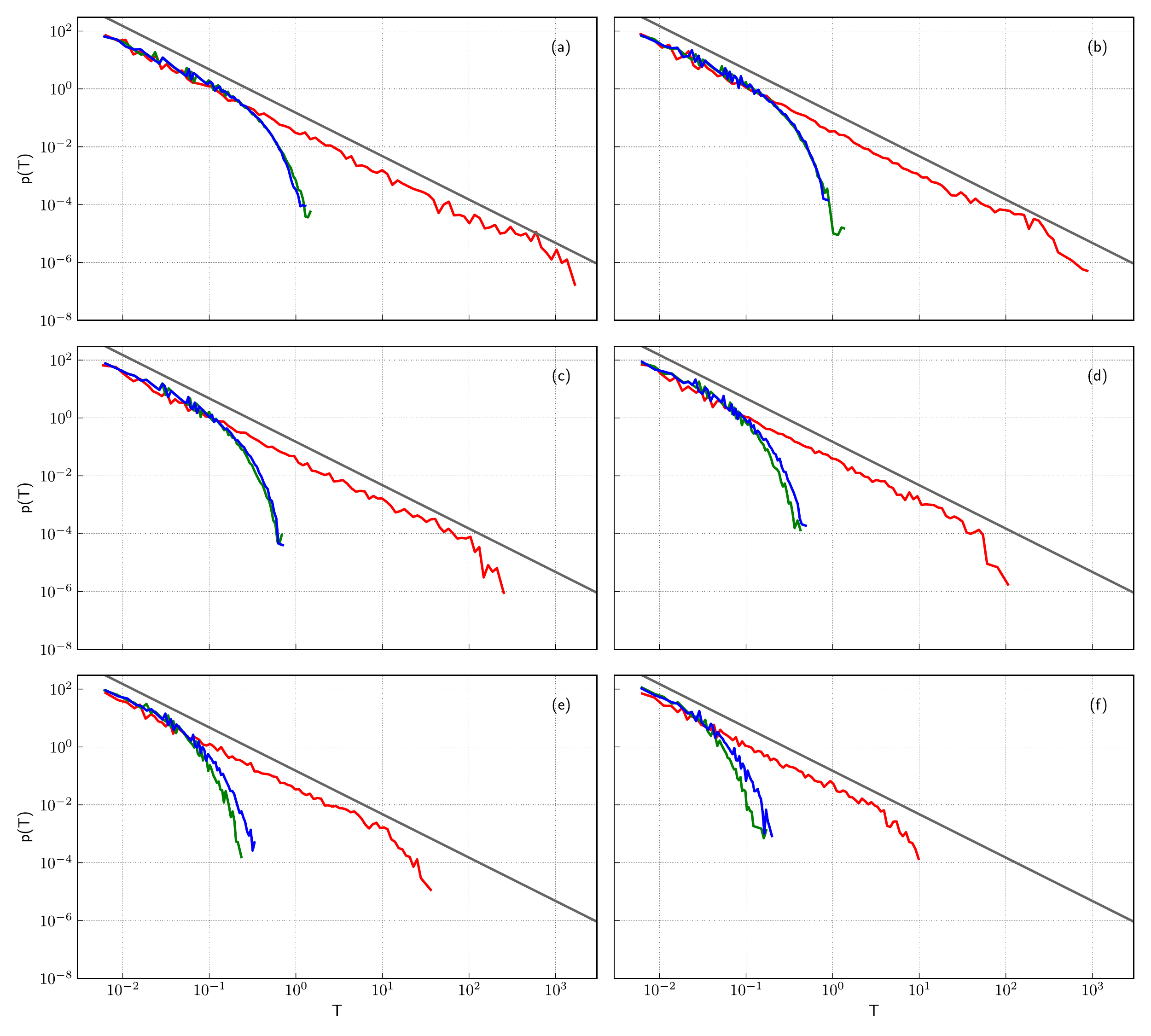}
\caption{\label{fig-3} PDFs of the burst duration $T$ for the model time series of absolute return calculated with various model compositions. $r_{\delta}(t)=y(t)$ series with (red curve), $r_{\delta}(t)=\vert y(t) \xi(t) \vert$ series (green curve), $r_{\delta}(t)=b_0(t)(1+a_0(\vert y(t) \xi(t) \vert)$ series (blue curve). Values of the threshold $q$ are the same as in Fig. \ref{fig-1}. The straight gray curves are shown to guide the eye showing a power-law with exponent $3/2$.}
\end{figure}

\begin{figure}
\centering
\includegraphics[width=\subfigfff]{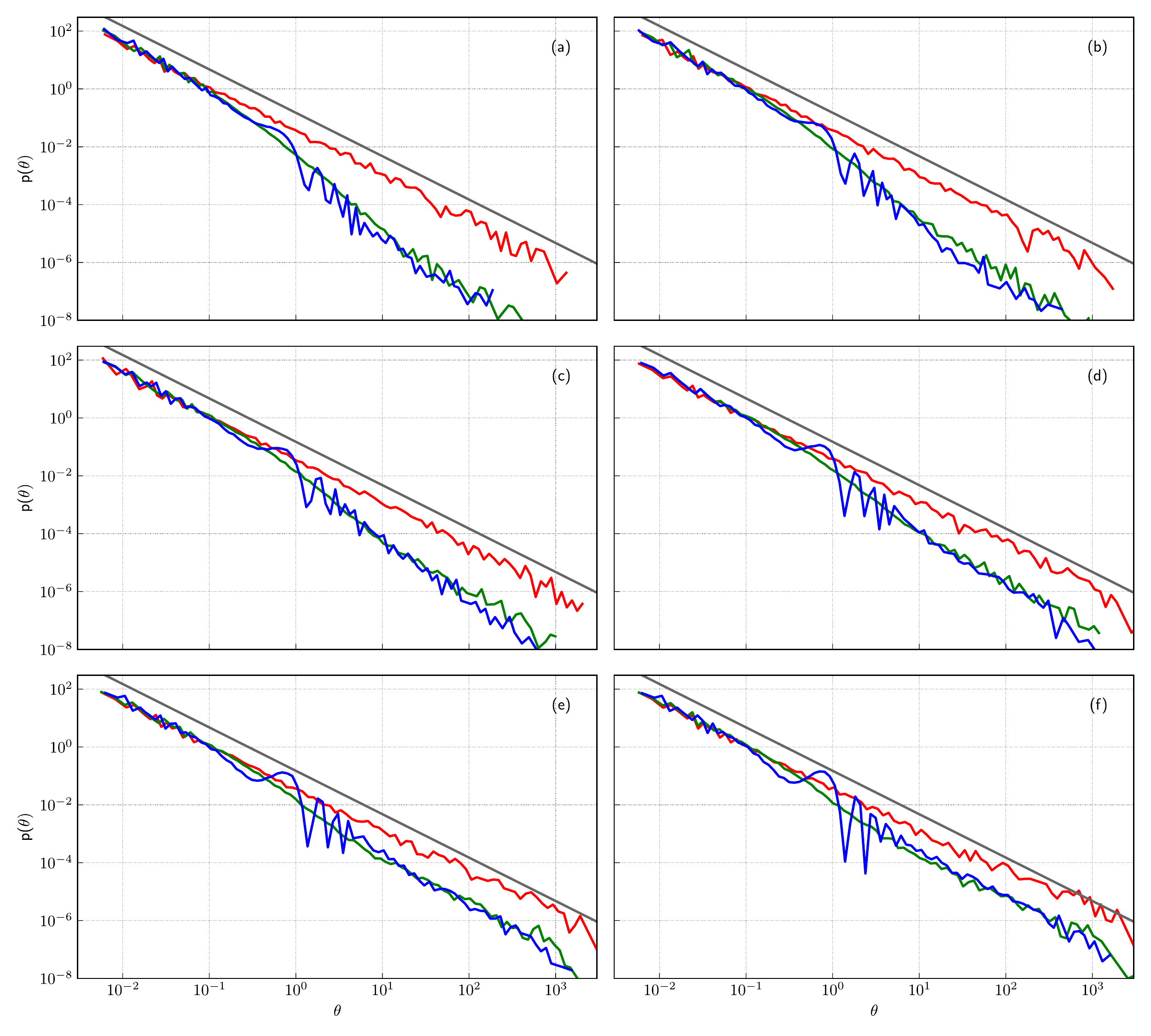}
\caption{\label{fig-4} PDFs of the inter-burst duration $\theta$ for the model time series of absolute return calculated with various model compositions. $r_{\delta}(t)=y(t)$ series with (red curve), $r_{\delta}(t)=\vert y(t) \xi(t) \vert$ series (green curve), $r_{\delta}(t)=b_0(t)(1+a_0(\vert y(t) \xi(t) \vert)$ series (blue curve). Values of the threshold $q$ are the same as in Fig. \ref{fig-1}. The straight gray curves are shown to guide the eye showing a power-law with exponent $3/2$.}
\end{figure}

Let us to simplify the consentaneous model by excluding all other noises and processes except the slowest one $y(t)$, which is defined by Eqs. (\ref{eq:nftau}) and (\ref{eq:taunfxi}) evolving according to the non-linear SDE similar to (\ref{eq:SDEy}), having the time scale parameter $h=0.3\cdot 10^{-8} \mathrm{s}^{-1}$ . Note here that real time $t$ is scaled in Eq.(\ref{eq:SDEy}) $t_s=h \cdot t$. Such model assumption would would mean the case, where return $r(t)$ is replaced by the average values in the subsequent time intervals $\delta$ of one-dimensional stochastic process $y(t)$. Here it is easy to predict the PDF of the burst and inter-burst duration, taking the power-law form with the exponent $3/2$ and exhibiting the exponential cutoff for extremely long durations. The explicit form of PDF for the burst duration with high threshold values was derived in \cite{Gontis2012ACS} and numerical analyses of burst and inter-burst duration for $y(t)$ defined by Eq. (\ref{eq:SDEy}) is given in \cite{Gontis2017Entropy}. From our point of view, the statistical properties of the return in the financial markets first of all are driven by the endogenous stochastic opinion dynamics described by the variable $y(t)$. In \figref{fig-3} and \figref{fig-4} we demonstrate the PDFs (red curves)  of the burst and inter-burst duration, accordingly, calculated numerically from Eqs. (\ref{eq:nftau}) and (\ref{eq:taunfxi}) for various values of thresholds. In \figref{fig-3} the observed cutoff of the power law $3/2$ for duration values $T \simeq 10^3 \mathrm{days}$ is in agreement with theoretical prediction in \cite{Gontis2012ACS}.

Next lets introduce the speculative dynamics of chartists modeled by the stochastic variable $\xi(t)$. Now the variable $\vert y(t) \xi(t) \vert$ becomes double stochastic, but still we neglect seasonality $b_0(t)$. PDFs of the burst and inter-burst duration in this case are given as the green curves in \figref{fig-3} and \figref{fig-4} accordingly. As one can observe in \figref{fig-4}, the double stochastic nature of the process increases the exponent of power-law in the region of the longer duration $\theta>1$. Obviously this could be confused with similar behavior of the true long-range memory processes with correlated increments having the exponent of PDF $\gamma=2-H$, when Hurst parameter is $H<0.5$. Probably this could be a problem when we have to decide from the empirical series which process is in the origin of the observed deviation from the main law of $3/2$. Fortunately for the intra-day time scales of duration $T$, $\theta<1$, where $\xi(t)$ dynamics dominates against $y(t)$,  the double stochastic process $\vert y(t)\xi(t)\vert$ exhibits power-law with exponent $3/2$ as well. PDF of $T$ in \figref{fig-3} exhibits only power-law  $3/2$ in the region $T\ll 1$ with subsequent exponential cutoff. 

Another component of the consentaneous model is related to daily seasonality introduced as periodic pattern in Eq. (\ref{eq:b0}). To account for the daily seasonality in FOREX we use here the width of intra-day pattern $w=0.25$, which is the only difference with parameter set used in \cite{Gontis2016PhysA}. We use the blue curves to plot the PDFs of the burst, \figref{fig-3}, and the inter-burst, \figref{fig-4}, duration in the return series $r(t)=b_0(t)(1+ a_0 \vert y(t) \xi(t) \vert)$ still keeping $\omega(t)\equiv 1$. In this very simplified version of seasonality we observe minor contribution to the statistical properties of burst and inter-burst duration as green and blue PDFs practically coincide, though some resonance structure can be observed in  \figref{fig-4}.  

\begin{figure}
\centering
\includegraphics[width=\subfigfff]{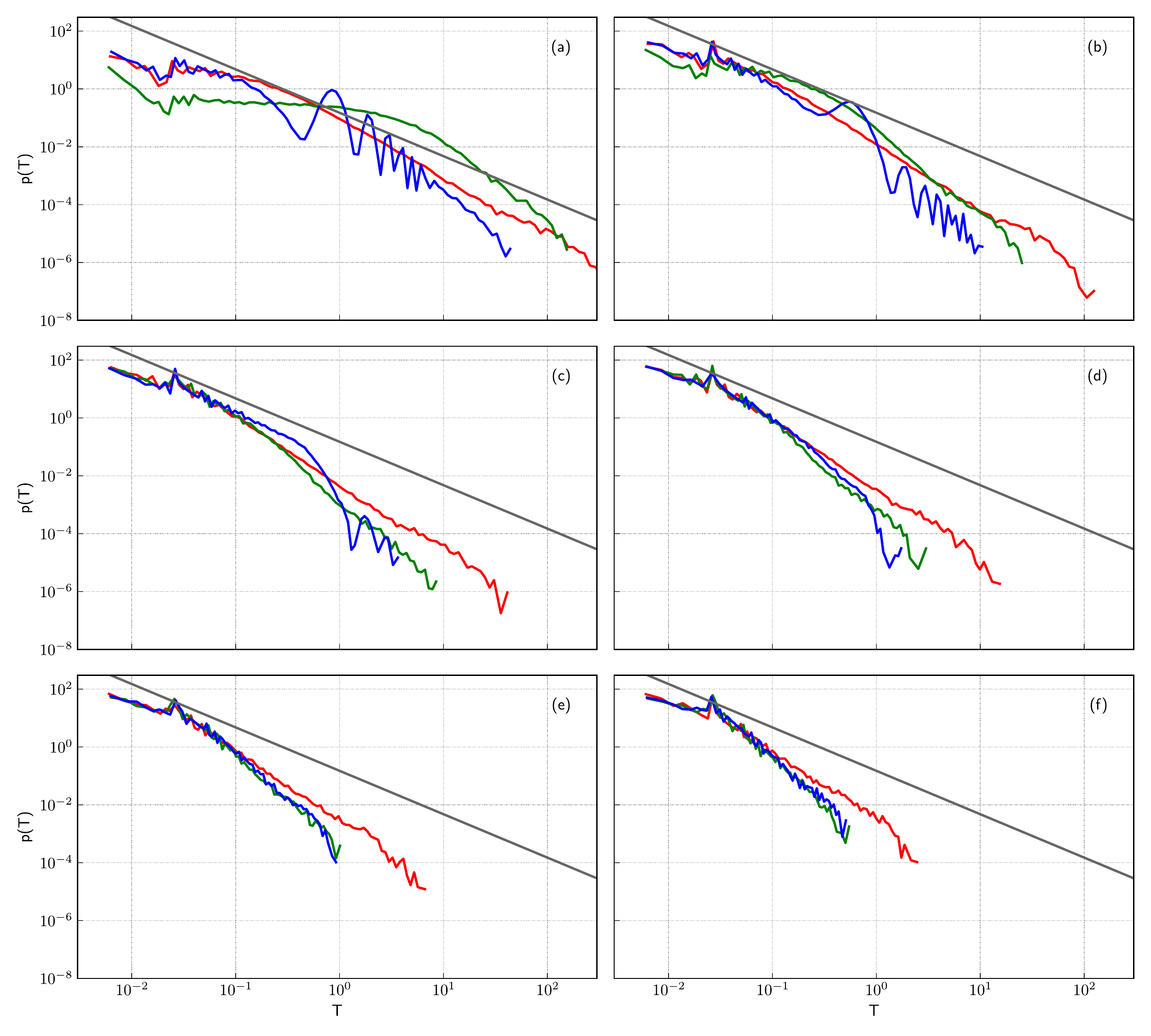}
\caption{\label{fig-5} PDFs of the burst duration $T$ for the model time series of the absolute return calculated accounting for the exogenous noise and with various other model components, Eqs. (\ref{eq:return}) and (\ref{eq:defvolatil}). $r_{\delta}(t)$ series   with $b_0(t)=\xi(t)\equiv 1$ in the model (red curve), $r_{\delta}(t)$ series with $b_0(t)\equiv 1$ (green curve), full scale $r_{\delta}(t)$ model (blue curve). Values of the threshold $q$ are the same as in Fig. \ref{fig-1}. The straight gray curves are shown to guide the eye showing a power-law with exponent 3/2.}
\end{figure}

\begin{figure}
\centering
\includegraphics[width=\subfigfff]{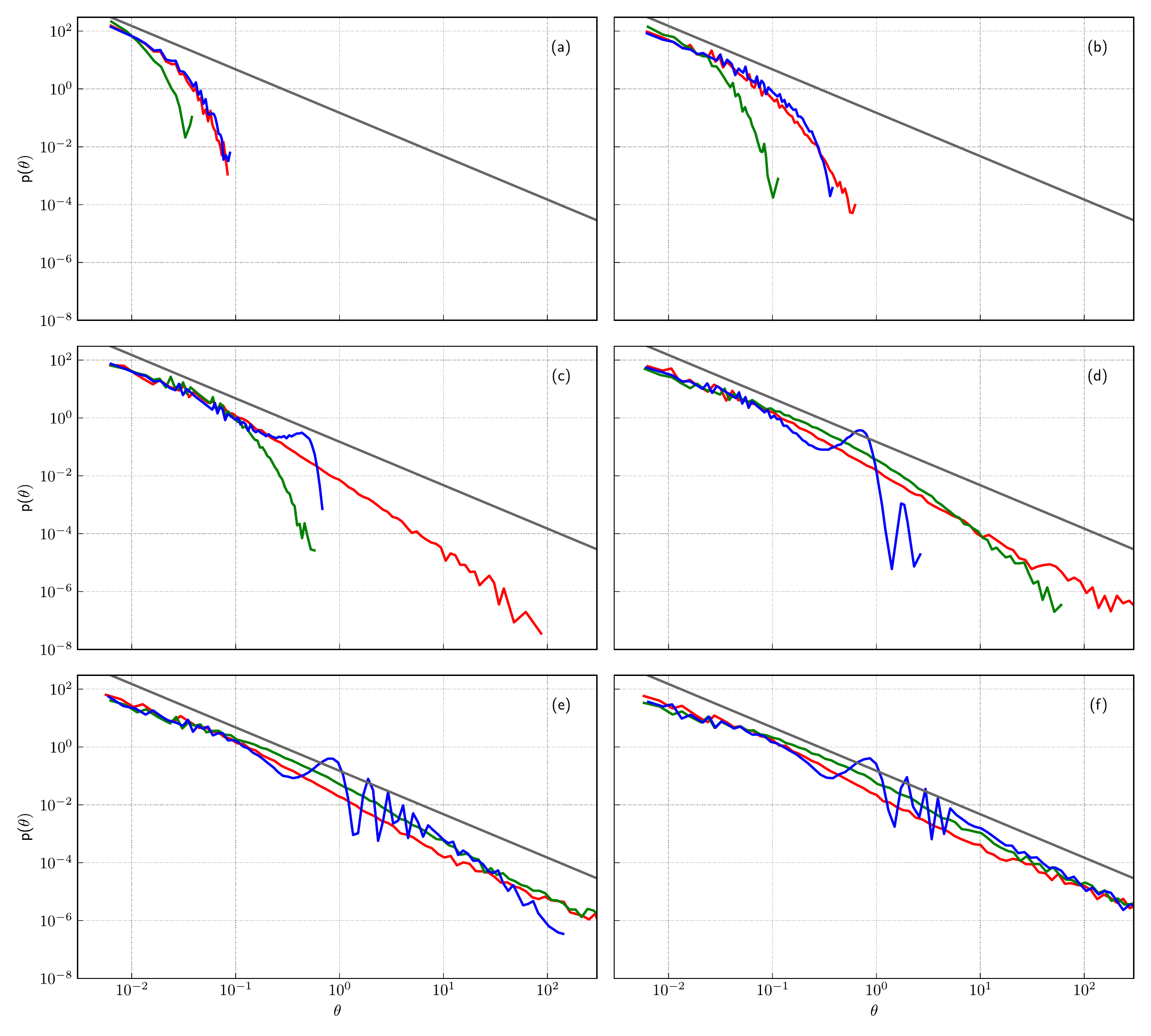}
\caption{\label{fig-6} PDFs of the inter-burst duration $\theta$ for the model time series of the absolute return calculated accounting for the exogenous noise and with various other model components, Eqs. (\ref{eq:return}) and (\ref{eq:defvolatil}). $r_{\delta}(t)$ series with $b_0(t)=\xi(t)\equiv 1$ in the model (red curve), $r_{\delta}(t)$ series with $b_0(t)\equiv 1$ (green curve), full scale $r_{\delta}(t)$ model (blue curve). Values of the threshold $q$ are the same as in Fig. \ref{fig-1}. The straight gray curves are shown to guide the eye showing a power-law with exponent 3/2.}
\end{figure}

In the full scale model, when we account for the exogenous noise $\omega(t)$ the interplay of all its components becomes more sophisticated. \figref{fig-5} and \figref{fig-6} are almost equivalent to \figref{fig-3} and \figref{fig-4}, where the meaning of the colors are the same with only the difference that now we switch the exogenous noise, $\omega(t)$, on and use standard deviation filter as in \figref{fig-1} and \figref{fig-2}. These figures are really illustrative showing us how the three independent Markov type noises and regular periodic fluctuation $b_0(t)$ interact generating complex enough behavior of burst and inter-burst duration PDF. 

The interaction of the one-dimensional Markov process $y(t)$ with exogenous Gaussian noise $\omega(t)$ is nicely reflected by red curves in \figref{fig-5} and \figref{fig-6}. The interplay of the two independent noises leads to the considerable deviation of the burst and inter-burst duration PDF from the power-law $3/2$. For the longer duration the exponent increases and for the short duration decreases. Note that there is no considerable region in \figref{fig-5} and \figref{fig-6}, where fundamental power-law is valid for red curve PDFs. This is qualitatively different behavior than in the full scale consentanous model and empirical analyzes \cite{Gontis2017PhysA}.

The deviations from power-law with exponent $3/2$ are more pronounced, when we add to the model one more independent Markov noise $\xi(t)$, green curves in \figref{fig-5} and \figref{fig-6}. The effect is really strong for the low values of threshold $q$. It would not be reasonable to expect power-law with exponent $3/2$ in the model if we were neglecting the regular pattern of seasonality. 

Fortunately the intra-day movement of the trading activity accounted by the regular pattern $b_0(t)$ in the model considerably changes the PDF of burst and inter-burst duration, see blue curves in \figref{fig-5} and \figref{fig-6}. Note, that these model PDFs are the same as black curves in \figref{fig-1} and \figref{fig-2}. Thus the full scale model, including all necessary noises of this complex system, is able to reproduce the empirical properties of the burst and inter-burst duration and explain the presence of considerable region in PDF with power-law exponent $3/2$. This serves as a strong argument that the absolute return fluctuations in the financial markets is a result of the interplay of at least three independent Markov noises and one regular movement of activity related with some intra-day pattern $b_0(t)$.

\section*{Conclusions}

The consentaneous model of the financial markets \cite{Gontis2014PlosOne} was proposed to reproduce PDF and PSD of the absolute return. Despite its capability to reproduce in details second order statistics usually considered as long-range memory of volatility, the model is memory-less as being based on agent and stochastic dynamics having Markov nature. Thus we consider the consentaneous model as exhibiting spurious nature of long-range memory arising from the interplay of independent endogenous and exogenous stochastic processes together with the regular intra-day movement of the trading activity. In our previous empirical study \cite{Gontis2017PhysA} we have concluded that observed power-law with exponent $3/2$ in PDF of burst and inter-burst duration confirms the possible spurious nature of long-range memory in finance. Nevertheless, this argument was not satisfactory, as exponent $3/2$ is a signature of Markov process only for one-dimensional stochastic fluctuations. The financial fluctuations are complicated enough as origin from endogenous and exogenous stochastic processes \cite{Gontis2016APPA}, thus the tests of long-range property  is not straightforward. 

Here we provided detailed numerical study of the consentaneous model analyzing its capability to reproduce the burst and inter-burst duration PDF of the absolute return in the FOREX. Our results confirm that the proposed model with the same set of the parameters is able to reproduce different statistical properties of the financial markets including PDFs of the burst and inter-burst duration for the various values of the threshold. Furthermore, with detailed analyzes how various noises interplay in the model we provided an evidence for all of its constituents: three state agent dynamics, exogenous noise and intra-day pattern, are important to get into agreement with empirical data. Our main conclusion in this contribution is that the consentaneous model is a strong argument for the spurious nature of long-range memory arising as a consequence of the non-linear stochastic dynamics. 

\singlespacing


\begin{thebibliography}{10}

\bibitem{Farmer2012EPJ}
J.~D. Farmer, M.~Gallegati, C.~Hommes, A.~Kirman, P.~Ormerod, S.~Cincotti,
  A.~Sanchez, D.~Helbing, A complex systems approach to constructing better
  models for managing financial markets and the economy, European Physics
  Journal Special Topics 214 (2012) 295--324.

\bibitem{Shiller2014AER}
R.~J. Shiller,
  \href{http://www.aeaweb.org/articles.php?doi=10.1257/aer.104.6.1486}{Speculative
  asset prices}, American Economic Review 104~(6) (2014) 1486--1517.

\bibitem{Kirman2014MD}
A.~Kirman, \href{http://journals.cambridge.org/article_S1365100514000339}{Ants
  and nonoptimal self-organization: Lessons for macroeconomics}, Macroeconomic
  Dynamics FirstView (2015) 1--21.

\bibitem{Kwapien2012PhysRep}
J.~Kwapien, S.~Drozdz, Physical approach to complex systems, Physics Reports
  515 (2012) 115--226.

\bibitem{Bouchaud2004Cambridge}
J.~P. Bouchaud, M.~Potters, Theory of financial risks and derivative pricing,
  Cambridge University Press, New York, 2004.

\bibitem{Sornette2004Princeton}
D.~Sornette, Why Stock Markets Crash: Critical Events in Complex Financial
  Systems, Princeton University Press, Princeton, USA, 2004.

\bibitem{Karsai2012NIH}
M.~Karsai, K.~Kaski, A.~L. Barabasi, J.~Kertesz, Universal features of
  correlated bursty behaviour, NIH Scientific Reports 2 (2012) 397.

\bibitem{Chakraborti2011RQUF1}
A.~Chakraborti, I.~M. Toke, M.~Patriarca, F.~Abergel, Econophysics review: I.
  empirical facts, Quantitative Finance 7 (2011) 991--1012.

\bibitem{Gabaix2009AR}
X.~Gabaix, Power laws in economics and finance, Annual Review of Economics 1
  (2009) 255--293.

\bibitem{Campbell1996Princeton}
J.~Campbell, A.~Wen-Chuan, A.~MacKinlay, The Econometrics of Financial Markets,
  Princeton University Press, Princeton, USA, 1996.

\bibitem{Mantegna2000Cambridge}
R.~N. Mantegna, H.~E. Stanley, Introduction to Econophysics: Correlations and
  Complexity in Finance, Cambridge University Press, 2000.

\bibitem{Drozdz2010NJP}
S.~Drozdz, J.~Kwapien, P.~Oswiecimka, R.~Rak, The foreign exchange market:
  return distributions, multifractality, anomalous multifractality and the epps
  effect, New J. Phys 12 (2010) 105003.

\bibitem{Brock2001RES}
W.~A. Brock, S.~N. Durlauf, Discrete choice with social interactions, Review of
  Economic Studies 68 (2001) 235--260.

\bibitem{Diks2005JEDC}
C.~Diks, R.~van~der Weide, Herding, a-synchronous updating and heterogeneity in
  memory in a cbs, Journal of Economic Dynamics \& Control 29 (2005) 741--763.

\bibitem{Franke2012JEDC}
F.~Franke, R.~Westerhoff, Structural stochastic volatiliy in asset pricing
  dynamics: Estimation and model contest, Journal of Economic Dynamics \&
  Control 36 (2012) 1193--1211.

\bibitem{Lux2012JEDC}
T.~Lux, Estimation of an agent-based model of investor sentiment formmation in
  financial markets, Journal of Economic Dynamics \& Control 36 (2012)
  1284--1302.

\bibitem{Goldbaum2014JEBO}
D.~Godlbaum, R.~C.~J. Zwinkels, An empirical examination of heterogeneity and
  switching in foreign exchange markets, Journal of Economic Behavior \&
  Organization 107 (2014) 667--684.

\bibitem{He2015JEF}
Y.~He, X.-Zh.~Li, Testing of a market fraction model and power-law behaviour in
  the dax 30, Journal of Empirical Finance 31 (2015) 1--17.

\bibitem{Jang2015CE}
T.-S. Jang, Identification of social interactions effects in financial data,
  Comput Econ 45 (2015) 207--238.

\bibitem{Gontis2010Intech}
V.~Gontis, J.~Ruseckas, A.~Kononovicius, A non-linear stochastic model of
  return in financial markets, in: C.~Myers (Ed.), Stochastic Control, InTech,
  2010, pp. 559--580.

\bibitem{Gontis2014PlosOne}
V.~Gontis, A.~Kononovicius,
  \href{http://journals.plos.org/plosone/article?id=10.1371/journal.pone.0102201}{Consentaneous
  agent-based and stochastic model of the financial markets}, PLoS ONE 9~(7)
  (2014) e102201.

\bibitem{Yamasaki2005PNAS}
K.~Yamasaki, L.~Muchnik, S.~Havlin, A.~Bunde, H.~Stanley, Scaling and memory in
  volatility return intervals in financial markets, Proceedings of the National
  Academy of Sciences of the United States of America 102 (2005) 9424--9428.

\bibitem{Wang2006PhysRevE}
F.~Wang, K.~Yamasaki, S.~Havlin, H.~Stanley, Scaling and memory of intraday
  volatility return intervals in stock market, Physical Review E 77 (2006)
  026117.

\bibitem{Wang2008PhysRevE}
F.~Wang, K.~Yamasaki, S.~Havlin, H.~Stanley, Indication of multiscaling in the
  volatility return intervals of stock markets, Physical Review E 77 (2008)
  016109.

\bibitem{Bunde2011EPL}
J.~Ludescher, C.~Tsallis, A.~Bunde, Universal behavior of the interoccurrence
  times between losses in financial markets: An analytical description, EPL 95
  (2011) 68002.

\bibitem{Bunde2014PRE}
J.~Ludescher, A.~Bunde, Universal behavior of the interoccurrence times between
  losses in financial markets: Independence of the time resolution, Physical
  Review E. 90 (2014) 062809.

\bibitem{Gontis2016PhysA}
V.~Gontis, S.~Havlin, A.~Kononovicius, B.~Podobnik, H.~Stanley, Stochastic
  model of financial markets reproducing scaling and memory in volatility
  return intervals, Physica A 462 (2016) 1091--1102.

\bibitem{Gontis2017PhysA}
V.~Gontis, A.~Kononovicius, Burst and inter-burst duration statistics as
  empirical test of long-range memory in the financial markets, Physica A 483
  (2017) 266--272.

\bibitem{Gontis2016APPA}
V.~Gontis,
  \href{http://przyrbwn.icm.edu.pl/APP/PDF/129/a129z5p24.pdf}{Interplay between
  endogenous and exogenous fluctuations in financial markets}, Acta Physica
  Polonica A 129~(5) (2016) 1023--1031.

\bibitem{Gontis2017Entropy}
V.~Gontis, A.~Kononovicius,
  \href{http://www.mdpi.com/1099-4300/19/8/387}{Spurious memory in
  non-equilibrium stochastic models of imitative behavior}, Entropy 19~(8)
  (2017) 387.

\bibitem{Rak2013APPB}
R.~Rak, S.~Drozdz, J.~Kwapien, P.~Oswiecimka, Stock returns versus trading
  volume: is the correspondence more general?, Acta Physika Polonica B 44
  (2013) 2035--2050.

\bibitem{Gabaix2003Nature}
X.~Gabaix, P.~Gopikrishnan, V.~Plerou, H.~E. Stanley, A theory of power law
  distributions in financial market fluctuations, Nature 423 (2003) 267--270.

\bibitem{Farmer2004QF}
J.~D. Farmer, L.~Gillemot, F.~Lillo, S.~Mike, A.~Sen, What really causes large
  price changes, Quantitative Finance 4 (2004) 383--397.

\bibitem{Gabaix2006QJE}
X.~Gabaix, P.~Gopikrishnan, V.~Plerou, H.~E. Stanley, Institutional investors
  and stock market volatility, The Quarterly Journal of Economics (May 2006)
  461--504.

\bibitem{Kaulakys2005PhysRevE}
B.~Kaulakys, V.~Gontis, M.~Alaburda, Point process model of 1/f noise vs a sum
  of lorentzians, Physical Review E 71~(051105) (2005) 1--11.

\bibitem{Ruseckas2011PhysRevE}
J.~Ruseckas, B.~Kaulakys, Tsallis distributions and 1/f noise from nonlinear
  stochastic differential equations, Physical Review E 84~(5) (2011) 051125.

\bibitem{Ruseckas2014JStatMech}
J.~Ruseckas, B.~Kaulakys, Scaling properties of signals as origin of 1/f noise,
  Journal of Statistical Mechanics (2014) P06004.

\bibitem{Ding1995fbm}
M.~Ding, W.~Yang, Distribution of the first return time in fractional brownian
  motion and its application to the study of on-off intermittency, Physical
  Review E 52 (1995) 207.

\bibitem{Gontis2012ACS}
V.~Gontis, A.~Kononovicius, S.~Reimann, The class of nonlinear stochastic
  models as a background for the bursty behavior in financial markets, Advances
  in Complex Systems 15~(supp01) (2012) 1250071.

\end{thebibliography}
\end{document}